# X-ray detection of SN1988Z with the ROSAT HRI

A.C. Fabian[1] and R. Terlevich[2]
1. Institute of Astronomy, Madingley Road, Cambridge CB3 0HA
2. Royal Greenwich Observatory, Madingley Road, Cambridge CB3 0EZ

**ABSTRACT**
We report the detection of SN1988Z in X-rays with the ROSAT High Resolution Imager. The inferred X-ray luminosity of about $10^{41}\,{\rm erg\,s^{-1}}$ can be well explained by assuming that the supernova occurred in a dense circumstellar medium, consistent with optical observations. SN1988Z is the most distant supernova yet detected in X-rays.

**Key words:**

## 1 INTRODUCTION

A rare subclass of Type II supernovae has been identified and associated with massive progenitor stars (Schlegel 1990; Filippenko 1989). These objects show narrow emission lines in addition to broad lines, lack P–Cygni profiles and reach high peak luminosities followed by a slower decay than that of normal Type II supernovae. Some are detected in the radio (e.g. SN1986J, Weiler et al 1990; SN1988Z, Van Dyk et al 1993) and X-ray bands (e.g., SN1986J, Bregman & Pildis 1992; SN1978K, Schlegel, Peter & Colbert 1995), and they tend to be associated with regions of active star formation. These supernovae probably occur in a dense circumstellar medium produced by mass-loss from the progenitor red giant (Terlevich 1994). Here we report the first detection of X-rays from SN1988Z, with a high bolometric X-ray luminosity of $\sim 10^{41}\,{\rm erg\,s^{-1}}$. At a redshift of $z = 0.022$, SN1988Z is the most distant supernova detected in X-rays.

The optical spectra and evolution of SN1988Z have been studied by Stathakis & Sadler (1991) and by Turatto et al (1993). The SN lies to the E of the nucleus of the spiral galaxy MCG+03 − 28 − 022 and has evolved very slowly, compared with more typical SNII. Radio observations of SN1988Z have been made by Van Dyk et al (1993), who note the strong similarity to SN1986J which led them to suggest that SN1988Z would be an attractive X-ray target.

Our X-ray observations are part of a programme to test the predictions of Terlevich et al. (1987, 1992) who assert that supernovae exploding in high density environments reach X-ray luminosities high enough to power the broad line region of some Active Galactic Nuclei (AGN). Here we present the X-ray detection of SN1988Z with the ROSAT High Resolution Imager (HRI).

## 2 THE ROSAT OBSERVATIONS

The region including SN1988Z was observed with ROSAT for 12287 s in the interval 1995 May 16–25. A point source with about 10 counts was detected by the HRI at the position (within the few arcsec pointing error) of SN1988Z $(10^{\rm h}51^{\rm m}50.0^{\rm s}, +16^{\rm d}00^{\rm m}1.7^{\rm s},$ J2000). The probability that this is a random fluctuation in the background is about $2 \times 10^{-5}$. Contours of the X-ray emission are shown on the digitized Sky Survey image in Fig. 1.

The hydrogen column density in our Galaxy along the line of sight to SN1988Z is about $3 \times 10^{20}\,{\rm cm^{-2}}$, which means that the (unabsorbed) source flux in the 0.2 – 2 keV band is about $3.5 \times 10^{-14}\,{\rm erg\,cm^{-2}\,s^{-1}}$. Assuming a 5 keV bremsstrahlung spectrum, the total source luminosity is then $(2 \pm 0.7) \times 10^{41}\,{\rm erg\,s^{-1}}$, if the temperature is 1 keV then the luminosity is $(1 \pm 0.3) \times 10^{41}\,{\rm erg\,s^{-1}}$ ($H_0 = 50\,{\rm km\,s^{-1}\,Mpc^{-1}}$). The error estimate is statistical and from the counts detected. It represents a minimum uncertainty given likely systematic uncertainties in the spectral shape.

## 3 DISCUSSION

Our detection of SN1988Z shows it to be a very luminous X-ray source. With a bolometric X-ray luminosity of at least $10^{41}\,{\rm erg\,s^{-1}}$ (unless it is all in line emission below 1 keV), the total energy radiated, if constant and isotropic, corresponds to about $2 \times 10^{49}$ erg or several percent of the total kinetic energy in the supernova.

The high optical and X-ray luminosity and the strong radio emission can all be the result of the interaction between the ejecta and a dense homogeneous circumstellar shell created by the slow wind from the progenitor (Chevalier 1982; Terlevich et al 1992; Chugai & Danziger 1994).

In this case there are 2 solutions for the observed X-ray luminosity; one for which the remnant is in an adiabatic Se-

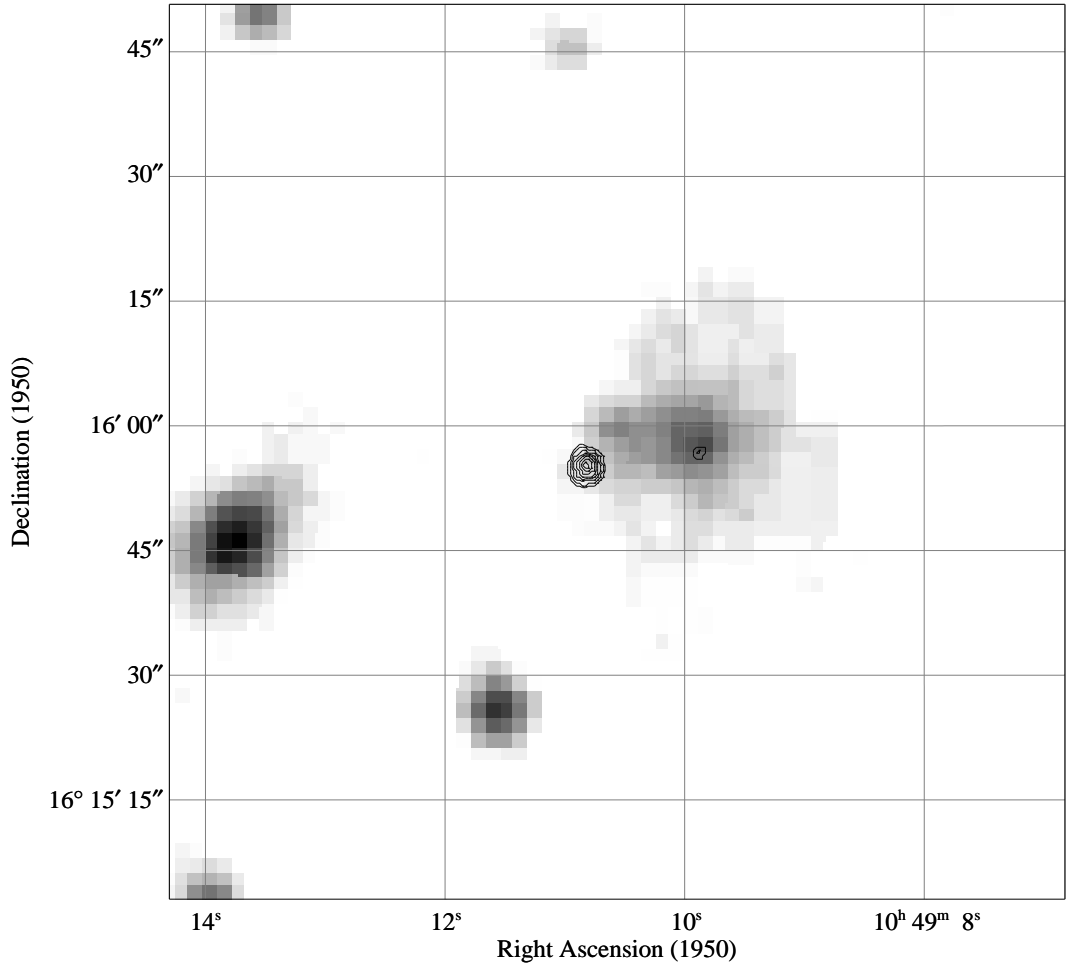

**Figure 1.** Contours of the ROSAT HRI image overlaid on the digitized UK Schmidt southern sky survey J plate. Note the good positional coincidence with SN1988Z, which lies 11 arcsec E and 2 arcsec S of the nucleus of the galaxy, the compact nature of the X-ray source, and the possible weak X-ray source ($\sim$ 30 per cent of the flux of the SNR) on the nucleus.

dov phase, the other in which it is in the radiative phase. A rough estimate of the SNR parameters for the adiabatic Sedov phase, assuming that the remnant is expanding into a medium of constant density $n$, can be obtained using the expressions given by Spitzer (1978). The bremsstrahlung luminosity is

$$L \approx 5 \times 10^{33} n^{6/5} t^{3/5} \text{ erg s}^{-1},$$

where the time is measured in years and the density in cm$^{-3}$. It is assumed that the X-ray emitting material is in the outer 10 per cent of the remnants radius, and that the density jump across the shock is the strong value of 4. Substituting the observed luminosity we find $n \approx 10^6$ cm$^{-3}$, the temperature is $T \approx 6 \times 10^7$ K and the radius is about $5 \times 10^{16}$ cm. The blast wave has swept up about 0.5 M$_\odot$ and the radiative cooling time of the post shock gas is about 52 yr, marginally long enough for the adiabatic assumption to be justified. This high value for the density is nevertheless an order of magnitude lower than the density measured for the circumstellar material emitting narrow forbidden lines,

$n \approx 10^7$ cm$^{-3}$ (Stathakis & Sadler 1991). Such material is of course at temperatures much less than that of the X-ray emitting plasma and so can be in denser embedded clouds.

Supernova remnants evolving in a dense, $n \gg 10^5$ cm$^{-3}$ and homogeneous medium, reach their maximum luminosity ($L > 10^7$ L$_\odot$) at small radii ($R < 0.1$ pc) soon after the SN explosion ($t < 20$ yr) while still expanding at velocities of more than 1000 km s$^{-1}$ (Shull 1980; Wheeler et al 1980; Draine and Woods 1991; Terlevich et al 1992; Terlevich 1994). In these compact SNR, radiative cooling becomes important well before the thermalization of the ejecta is complete. As a result, the Sedov phase is avoided and the remnant goes directly from the free expansion phase to the radiative phase. The shocked matter undergoes a rapid condensation behind both the leading and the reverse shocks. As a consequence two high-density, fast-moving thin shells are formed. These dense shells, the freely expanding ejecta and a section of the still dynamically unperturbed interstellar gas, are all ionized by the radiation from the shocks. The emitted spectrum resembles that of the broad-line region of AGN (Terlevich et al 1992).

Terlevich (1994) has shown that the simple homogeneous CSM model of compact SNR gives a reasonable description of the observed H$\alpha$ light curve of SN1988Z both in amplitude and time scale (see also Tenorio-Tagle et al. 1995). The same model predicts that a compact SNR evolving in a medium of constant density $n = n_7 10^7$ cm$^{-3}$, with $n_7 \sim 1$, will reach a peak luminosity of $2 \times 10^{43}$ erg s$^{-1}$ and temperature $3 \times 10^8$ K $\approx$ 30 keV corresponding to a shock velocity of about 5000 km s$^{-1}$ about eight months after the SN explosion. After the maximum the luminosity will decay as

$$L \approx 9 \times 10^{42} n_7^{-3/7} t^{-11/7} \text{ erg s}^{-1},$$

and the shock temperature as

$$T \approx 10^8 n_7^{-6/7} t^{-10/7} \text{ K},$$

and shock velocity

$$V_{\text{sh}} \approx 3400 t^{-5/7} \text{ km s}^{-1}.$$

Taking as the epoch of the maximum, 1988 December 1 (Stathakis & Sadler 1991; Turatto et al 1993), the ROSAT observations correspond to an age of 6.5 yr. The simple homogeneous CSM model predicts for $n = 10^7$ cm$^{-3}$ that 6.5 yr after reaching its maximum light the compact SNR will have a total shock luminosity of $4 \times 10^{41}$ erg s$^{-1}$ with a temperature $T \approx 10^7$ K $\approx$ 1 keV, shock velocity $\approx$900 km s$^{-1}$ and radius $\approx 6 \times 10^{16}$ cm. Only half of the shock luminosity, i.e. $2 \times 10^{41}$ erg s$^{-1}$, would be emitted outwards and not reprocessed by the cold dense shells. The predicted H$\alpha$ luminosity of the shells is $2 \times 10^{40}$ erg s$^{-1}$. The leading shock has swept about 9 M$_\odot$ and has a cooling time of less than one month.

The above X-ray luminosity is within a factor of two of the observed value (Fig. 2; using the lower temperature estimate from Section 2). This is in remarkable agreement given that there may be some absorption local to the SN and line emission has not been included in the estimate. Also the CSM may not be of constant density but decrease outward in some manner dependent of the progenitor star. We have adopted a constant density above for simplicity. It must terminate at a radius within a factor of 2 of the present one or the total mass would be prohibitive, the maximum age of the X-ray luminous phase in both solutions would then be about 30 y.

On the issue of whether such SNe can power AGN, we note on the positive side that some SNe can be efficient and luminous X-ray emitters and have AGN-like optical spectra as predicted by Terlevich and collaborators (Terlevich et al 1992; Terlevich 1994). On the negative side we note that a rather large number of such (rare) SNe (i.e. similar to 1988Z 6.5 yr after explosion) would be required to provide the broad line region of a typical Seyfert 1 galaxy (at say $10^{43}$ erg s$^{-1}$). The conditions are relaxed if the SNe are much more luminous in the first month or so, as optical observations and models would indicate. Even then the present understanding of the evolution of SN into uniform CSM does not predict the observed large amplitude X-ray variability on hours to minutes often seen in such AGN. Further X-ray observations of such SNe when very young are required.

Alternatively the power and X-ray luminosity could be due to a young pulsar. Scaling from the Crab pulsar, we find that for a power of $10^{41}$ erg s$^{-1}$ the spin period has to be about 10 ms. The power then declines with time as $t^{-2}$. Such a pulsar would need to be extraordinarily efficient in X-ray production to account for the observed emission.

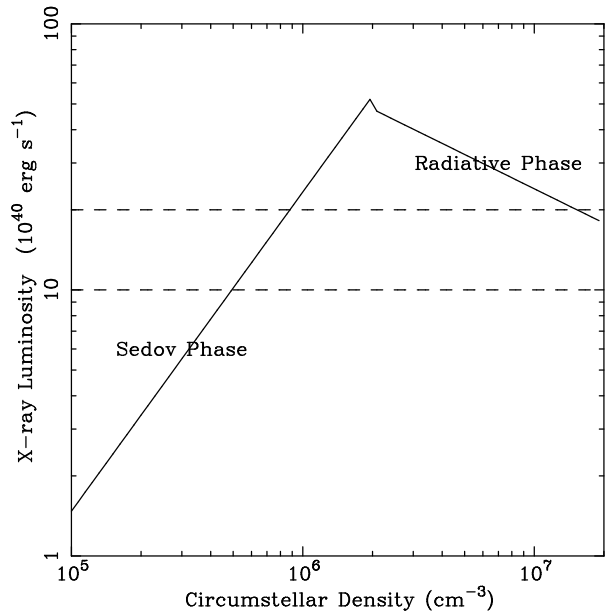

**Figure 2.** Schematic illustration of the two density solutions for the observed X-ray luminosity at $t = 6.5$ yr. The luminosity for the radiative phase has been reduced by a factor of 2 in order to compare with the outward observable luminosity. The horizontal dashed lines represent the inferred X-ray luminosity of SN1988Z for an assumed temperature of 5 keV (upper line) and 1 keV (lower line). Note that the shock temperature decreases below the observable ROSAT band if the density much exceeds a few times $10^7$ cm$^{-3}$.

As a final comment, we note that some of the very luminous X-ray sources (with $L \sim 10^{40}$ erg s$^{-1}$) seen in normal galaxies (see e.g. Fabbiano 1988) could be cSNR similar to, but older than, SN1988Z. If the circumstellar density is slightly lower than is the case for SN1988Z, then they can last longer. The identification of such objects, perhaps through radio emission, could help to estimate the frequency of such supernovae.

## 4 SUMMARY

SN1988Z has given rise to an X-ray luminous SNR. This is plausibly due to the SN occurring in a dense surrounding medium. Assuming that the medium is of constant density, the SNR can either be in a Sedov phase with the density at $\sim 10^6$ cm$^{-3}$ or near the end of a strongly radiative phase. These possibilities can be distinguished by future observations of the rate of decay of the X-ray flux and spectrum.

**ACKNOWLEDGEMENTS**

ACF thanks the Royal Society for support. RJT thanks ESO for hospitality.

**REFERENCES**

Bregman J.N., Pildis R.A., 1992, ApJ, 398, L107
Chevalier, R.A., 1982, ApJ, 259, 302
Chugai, N.N. & Danziger, I.J., 1994, MNRAS, 268, 173
Draine, B.T. & Woods, D.T., 1991, ApJ, 383, 621
Fabbiano G., 1989, ARAA, 27, 87
Filippenko, A., 1989, AJ, 97, 726
Schlegel E.M., 1990, MNRAS, 244, 269
Schlegel E.M., Petre R., Colbert E.J.M., 1995, ApJ submitted
Spitzer L., 1978, Physical Processes in the Interstellar Medium, Wiley
Shull, J.M., 1980, ApJ, 237, 769
Stathakis R.A., Sadler E.M., 1991, MNRAS, 250, 786
Tenorio-Tagle, G, Terlevich, R., Rozyczka, M. & Franco, J., 1995, Submitted to MNRAS
Terlevich, R., Melnick, J. & Moles, M., 1987, Proceedings of the IAU Symposium 121, ed. Khachikian et al., Reidel; p 499
Terlevich, R., 1994, Proceedings of the 34th Herstmonceux Conference, Cambridge University Press; p 153
Terlevich, R., Tenorio-Tagle, G, Franco, J. & Melnick, J., 1992 MNRAS, 255, 713
Turatto M., Cappellaro E., Danziger I.J., Benetti S., Gouiffes C., Della Valle M., 1993, MNRAS, 262, 128
Weiler, K.W., Panagia, N. & Sramek, R.A., 1990, ApJ, 364, 611
Wheeler, J.C., Mazurek, T.J. & Sivaramakrishnan, A., 1980, ApJ, 237, 781
Van Dyk S., Weiler K., Sramek R.A., Panagia N., 1993, ApJ, 419, L69